%% file: main.tex
\documentclass[runningheads]{llncs}
\usepackage{graphicx}
\usepackage[figuresright]{rotating}
\usepackage[table]{xcolor}
\definecolor{mywhite}{RGB}{255, 255, 255}
\definecolor{mygray}{RGB}{240, 240, 240}
\usepackage{tabularx}
\usepackage{supertabular}
\usepackage{longtable}
\usepackage{lscape}
\usepackage{ltablex}
\usepackage{ctable}
\usepackage{multirow}

\usepackage{hyperref}
\hypersetup{
    colorlinks=false,
    linkcolor=blue,
    filecolor=magenta,      
    urlcolor=cyan,
}

\usepackage{floatrow}
\input{latexTools.tex}

\begin{document}
\title{Training Medical Communication Skills with Virtual Patients: Literature Review and Directions for Future Research}
\titlerunning{Training Medical Communication Skills with VPs}
%

\author{Edoardo Battegazzorre\orcidID{0000-0002-1383-3285} \and
Andrea Bottino\orcidID{0000-0002-8894-5089} \and
Fabrizio Lamberti\orcidID{0000-0001-7703-1372}}

%
%

\institute{DAUIN, Politecnico di Torino, Corso Duca degli Abruzzi 24, 10143 Torino, Italy 
\email{\{edoardo.battegazzorre, andrea.bottino, fabrizio.lamberti\}@polito.it}}

%
\maketitle              
\input{abstract}

\input{introduction}

\input{searchProtocol}

\input{taxonomy}


\input{openResearch}

\input{conclusion}



%
%
%
\bibliographystyle{splncs04}

\end{document}

%% file: latexTools.tex
\usepackage{xcolor} 
\usepackage{soul} 
\usepackage{xspace}
\usepackage{amsmath}
\usepackage{multirow}
\usepackage{capt-of}
\usepackage{floatrow}

\usepackage{hyphenat}
\newcommand{\totalArticles}{28\xspace} 

\newcommand{\totalVPs}{21\xspace} 

\newcommand{\quotes}[1]{``#1''}

%% file: abstract.tex
\begin{abstract}

Effective communication is a crucial skill for healthcare provi-ders 
since it leads to better patient health, satisfaction and avoids malpractice claims. In standard medical education, students' communication skills are trained with role-playing and Standardized Patients (SPs), i.e., actors. However, SPs are difficult to standardize, and are very resource consuming. Virtual Patients (VPs) are interactive computer-based systems that represent a valuable alternative to SPs. VPs are capable of portraying patients in realistic clinical scenarios and engage learners in realistic conversations. Approaching medical communication skill training with VPs has been an active research area in the last ten years. As a result, the number of works in this field has grown significantly. The objective of this work is to survey the recent literature, assessing the state of the art of this technology with a specific focus on the instructional and technical design of VP simulations. After having classified and analysed the VPs selected for our research, we identified several areas that require further investigation, and we drafted practical recommendations for VP developers on design aspects that, based on our findings, are pivotal to create novel and effective VP simulations or improve existing ones.


\end{abstract}

%% file: introduction.tex
\section{Introduction}
\label{sec:intro}

The communication between patients and doctors is a central component of health care practice. On the one hand, good doctor-patient communication help physicians better identify patient's needs, perceptions and expectations \cite{Longnecker2010}.
On the other hand, it is not surprising that patients rate open communication as one of the most important aspects of their relationship with the physicians \cite{dibbelt2010patient}.
Research has shown that an effective, patient-centered communication is important to increase patient satisfaction  \cite{fiscella2004patient,franks2005patients,hickson2002patient,kee2018communication,kohatsu2004characteristics,papadakis2005disciplinary,stelfox2005relation}. It can have also beneficial effects on patient's health, improving physiologic measures as blood pressure and glucose levels \cite{stewart1995effective}, increasing understanding and adherence to therapy \cite{King2013}, and even creating a placebo effect in some cases \cite{kelley2009patient}.  Conversely, a poor grasp of communication skills can be detrimental to both the patient's and their relatives' health \cite{judge2004affect,kee2018communication}, and may lead to malpractice accusations.

Communication is a complex phenomenon that's not restricted to the verbal domain. As outlined in \cite{kee2018communication}, there are several critical aspects that have been subject to patients' complaints. They include elements of non-verbal communication (e.g., lack of eye-contact, negative facial expressions, use of \quotes{improper} prosodic features), inappropriate choice of words, lack of pauses to let patients ask questions, lack of listening, issues with the information given, and lack of empathy or even disrespect and poor attitudes.

Given the relevance of the problem, a primary goal in healthcare is training clinicians and healthcare providers in developing effective communication skills. Today, standardized patients (SP, i.e., actors who are instructed to represent a patient during a clinical encounter with a healthcare provider) are considered the gold standard in such training programs. SPs provide students with the opportunity to learn and practice both technical and non-technical skills in an environment capable of reproducing the realism of the doctor-patient relationship. These simulated environments are less stressful for the students, who are not required to interact with a real patient \cite{forrest2013essential,kneebone2006human}. 
However, SPs are difficult to standardize, since their performance heavily depend on the actors' skills, and their recruitment and training can become very costly \cite{Nestel2011}.

A practical alternative to SPs is represented by virtual patients (VPs), i.e., interactive computer simulations capable of portraying a patient in a clinical scenario in a realistic way. VPs are virtual agents that have a human appearance and the ability to respond to users and engage in communication patterns typical of a real conversation. They can be equipped with external sensors capable of capturing a wide range of non-verbal clues (user's gestures and motions, expressions and line of sight) and use them to modulate the evolution of the conversation. In the field of communication skill learning, VPs have the same advantages over SPs and are characterized by comparable learning outcomes \cite{quail2016student}. They are cost-effective solutions since they can be developed once and used many times. They can be deployed as in-class or self-learning tools that students can use at their own pace and at any place. They can also be integrated into sophisticated software platforms that include automatic learner assessment, feedback and debriefing sessions. Finally, VP simulations can present, with reasonable accuracy, difficult or rare cases with a high degree of repeatability \cite{urresti2017virtual} and, compared to SPs, are also easier to standardize \cite{rogers2011developing}. 

VPs for provider-patient communication have been surveyed in several works. Bearman et al. \cite{bearman2015learning} conducted a systematic review on VPs focused on developing empathy-related skills, but the Authors did not extend their analysis to communication as a whole. A more recent review \cite{richardson2019virtualreview} investigated the specific context of pharmacist-patient counselling, focusing on the development of knowledge, skills, confidence, engagement with learning, and user satisfaction. However, the work did not discuss in depth the effects of specific design choices. Another integrative review \cite{peddle2016virtual} focused on non-technical skills like situational awareness, decision-making, teamwork, leadership, and communication, but did not consider the technical perspective and, like the previous one, it did not elaborate on the impact of specific instructional or design features.
Finally, the Authors of \cite{lee2020effective} performed a systematic review on VPs focused on communication, analyzing which features of instructional design (i.e., the definition of methods, processes and strategies that guide learners to achieve the training objectives) and technical design (i.e., the definition of the technological components aimed to support and implement the envisaged instructional design) are most effective in VP simulations. Unfortunately, the time span of the survey was limited to 2006--2018, and the number of studies remaining from the application of the inclusion and exclusion criteria was rather small (14 works, with only eight discussed in detail). 

Based on the above analysis, it was our opinion that a thorough analysis of the instructional and technical design elements as well as of the technological components (sensory system, speech understanding, interaction devices, virtual reality, VR, and augmented reality, AR, etc.) and related relevant concepts (like immersion and presence) that are involved in the development of the considered learning tools was actually missing. Hence, we tried to fill this gap through the review reported in the present paper. Similarly to \cite{lee2020effective}, our study is centered on the instructional and technical design of  VP simulations. However,  we propose a different approach to the analysis of these two components, by performing in particular a deeper investigation of the technical aspects. Then, based on our findings, we also identify current limitations and potentially unexplored areas with the aim to foster further research and developments in the field.

The rest of the paper is organized as follows. In Section \ref{sec:reviewProtocol}, the literature review protocol is first introduced. Afterwards, Section \ref{sec:taxonomy} presents and discusses the research results.  Section \ref{sec:openResearch}  highlights the gaps that we identified  and the directions that future studies could take in order to address
them. Finally, conclusions are given in Section \ref{sec:conclusion}.

%% file: searchProtocol.tex
\section{Literature review protocol}
\label{sec:reviewProtocol}

As illustrated in the previous section, we performed a literature review to document the current state-of-the-art of the use of VPs for medical communication skill learning and to identify possible areas where further research is needed. The purpose of this review was to understand the instructional and technical design principles and the efficacy of these elements in achieving the expected learning outcomes.
To this end, we developed the following guiding questions in order to help focusing information extraction.

\begin{itemize}
\item RQ1: What are the latest technical developments in the field of communica-tion-oriented VPs? 

\item RQ2: Which instructional and technical design features are employed most commonly in VP design?

\item RQ3: Which instructional and technical VP design features are more effective for learning communication skills? And which of these features are most appreciated by the users?


\end{itemize}

The search process, carried out mainly between March and May 2020, started with an automated approach targeting four scientific paper databases, namely Scopus, PubMed, ACM Digital Library and IEEE Xplore. For each database we performed a search based on the main and derivative keywords (virtual patient OR (serious game AND healthcare)) AND (communication), limiting the results to papers published from 2015 onward. The choice of this date was made with the aim to survey only the most recent developments in the field and avoid excessive overlaps with previous literature reviews (e.g., with \cite{peddle2016virtual} and \cite{lee2020effective}). 

The papers found were post-processed in order to remove repeated entries and exclude reviews, editorials, abstracts, posters and panel discussions. The remaining 306 papers were analyzed by reading over their title, abstract, and introduction, and classified as either relevant or irrelevant based on the following criteria: (i) does the study relate to any of the design elements of interest (instructional or technical)? and, (ii) does the study disclose at least some of the design choices made by the Authors? If the answer to any of these questions was no, then the paper was excluded. After this step, each of the 70 accepted papers was read completely by at least one reviewer, who also assessed its quality. Its references were also analyzed according to the aforementioned screening process. 

At the end of the search process, we selected a total of \totalArticles papers. Among them, we identified a number of works that referred to the same VP, but in different experimental settings or in different phases of the development process. Since our interest was in analyzing the VP design rather than the detailed outcomes of possible experiments, papers sharing the same VP were grouped together, obtaining a total of \totalVPs VPs (17 of them had not been discussed in previous surveys, and only four of them were in common with \cite{lee2020effective}, namely Banszki \cite{banszki2018clinical,quail2016student}, CynthiaYoungVP \cite{foster2016using}, MPathic-VR \cite{guetterman2019medical,kron2017using}, and NERVE \cite{hirumi2016advancingPart2,hirumi2016advancing,kleinsmith2015understanding}). 

In order to capture the main characteristics of problems and solutions discussed in these papers, we introduced a taxonomy of terms for the instructional and technical design elements, whose initial version was defined based on the Authors’ expertise. Based on intermediate findings, this taxonomy was further refined into the final one introduced in Section~\ref{sec:taxonomy}. All the Authors categorized the selected VPs according to this taxonomy, and any disagreement was solved by discussion.
Finally, as a last step, we searched for references related to the open problems and potential areas of research identified during the analysis.

%% file: taxonomy.tex
\section{Results and discussion}
\label{sec:taxonomy}

In this section, we present the result of our research. As stated in Section \ref{sec:reviewProtocol}, the selected VPs have been labeled according to the taxonomy of terms summarized in Tables \ref{table:instructionalDesignTable} and \ref{table:technicalDesignTable}. The definition of the identified categories (which differs to a large extent from the one presented in \cite{lee2020effective}) is introduced in the following subsections, where we also discuss the survey results relative to each group. We first introduce the instructional design elements, which are connected to the technical elements necessary to realize them; afterwards, we discuss the design choices related to the technical and technological components of the simulations. Finally, we discuss the experimental evidences related to the effectiveness of the identified design elements.




\subsection{Instructional design}
\label{sec:instructionalDesign}

This category encompasses various instructional design aspects implemented in the VP scenario, such as how the VP delivers (and facilitates) learning activities and if (and how) it provides scaffolded support to improve learner's performance.

\input{tables/instructionalTable}

\input{tables/technicalTable}

\textbf{Structure}. The \emph{structure} defines the hierarchical organization and presentation of VP-related information within the simulation.
According to \cite{bearman2001random}, two non-mutually exclusive approaches (i.e., \emph{narrative} and \emph{problem solving}) can be defined. 
The narrative VPs are characterized by a coherent storyline, with a focus on cause-effect decisions that have a direct impact on the evolution of the simulation. These VPs present the patient as more than a mere collection of data and statistics, and devote a certain degree of attention to interpersonal and communication aspects of the provider-patient interaction.  On the contrary, the \emph{problem solving} VPs are mainly used to support inquiry-based learning scenarios such as teaching clinical reasoning, differential diagnosis, and history-taking skills. These contexts do not usually concern  portraying authentic communicative acts, since they mainly involve making questions and observations. 

Scholars and researchers recognize the power of \emph{narrative} design in the creation of meaningful learning experiences \cite{bearman2001random,marei2018use}.
Narrative-based simulations that reflect the consequences of the choices and the actions made by the learner can lead to the development of more effective VPs.  In particular, for VPs used to teach communication skills, experimental  evidence supports the value of \emph{narrative} design \cite{bearman2001random}. Thus, it is not surprising that all the VPs presented in the selected works are based on this approach. Nevertheless, it is interesting to note that 10 out the \totalVPs VPs analyzed integrate the \emph{narrative} design with a \emph{problem solving} component. 
This component aims to teach particular skills like history-taking (Cynthia Young VP \cite{foster2016using}, NERVE \cite{hirumi2016advancingPart2,hirumi2016advancing,kleinsmith2015understanding}, Maicher \cite{maicher2017developing}), clinical reasoning (VSPR \cite{peddle2019exploring,peddle2019development}, Richardson \cite{richardson2019virtual},  Washburn \cite{washburn2020virtual}, UT-Time Portal \cite{zielke2016beyond,zielke2016using}, Zlotos \cite{zlotos2016scenario}),  physical examinations (HOLLIE \cite{adefila2020students}, NERVE \cite{hirumi2016advancingPart2,hirumi2016advancing,kleinsmith2015understanding},  CESTOL VR Clinic \cite{sapkaroski2018implementation}), compilation and consultation of electronic medical records (HOLLIE \cite{adefila2020students}, Maicher \cite{maicher2017developing}), and medication administration (HOLLIE \cite{adefila2020students}).

\textbf{Unfolding}. Given the prominence of narrative design in the development of VPs for communication skill training, another relevant design element is defining how the narrative may unfold, and how the simulation can evolve between different states. A preliminary subdivision can be made among \textit{linear} and \textit{non-linear} narratives. In the former design, VPs have a linear path to follow and the decisions, questions and options possibly presented to the learner do not influence the simulation outcome. It is clear that this design severely limits learning effectiveness, and none of the works included in this survey implemented it.

On the contrary, the \emph{non-linear} navigational structure of VPs offers learners a greater flexibility, and an higher degrees of interactivity and control. In this case, two further choices are possible. In the \textit{closed-option} design, the simulation advances to the next state by selecting one of the possible alternatives or explicit paths offered to learners. Simulation states are organized in a hierarchical structure (similar to that of the \quotes{choose your own adventure} books), which stresses the cause-effect relation of the user's choices. 
The \textit{open-option} design (sometimes referred in the literature as \quotes{free-text} \cite{jacklin2019virtual,janda2004simulation,mccoy2016evaluating}  or \quotes{open-chat} \cite{hirumi2016advancing}) can be used to develop  free-form simulations where states are organized in a partially or fully interconnected structure, and users are free to interact with the VP as they wish, thus emulating the flow of a real conversation. As we will discuss  more in detail in Section \ref{sec:technicalDesign}, learners can formulate questions and statements by either typing or having their speech transcribed into written words using speech-to-text software. Then, the application parses the text and elaborates a proper response. The VP state progression can be influenced also by non-verbal cues such as gestures, body posture, expressions and sight.


The \textit{closed-option} design characterizes most of the analyzed VPs (15), with only four works  based on an \textit{open-option} design; as for the remaining, one VP implemented both options (NERVE \cite{hirumi2016advancingPart2,hirumi2016advancing,kleinsmith2015understanding}), whereas the other one can be considered an hybrid between the two designs (Ochs \cite{ochs2019training}). One explanation for this result is the lower complexity of the \textit{closed-option} implementation, although some Authors \cite{carnell2015adapting,jacklin2019virtual} argued that a such an approach may be more suitable for novices who, for example, may still be inexperienced about the procedures to follow in a patient encounter. However, other works reported that many students feel restricted by the \textit{closed-option} interface \cite{dupuy2019virtual,hirumi2016advancing,jacklin2019virtual,peddle2019development}, preferring either an \textit{open-option} structure or the possibility to chose between the two variants. It is worth noting that implementing both options allows the use of the same VP in different educational settings. 
For instance, in NERVE \cite{hirumi2016advancingPart2,hirumi2016advancing,kleinsmith2015understanding}, the less stress-inducing \textit{closed-option} variant is used in the learning sessions, while rehearsal sessions leverage the less restrictive \textit{open-option} setting. 
Another interesting approach is the hybrid model implemented in Ochs \cite{ochs2019training}, where the user can freely interact through voice with the VP. Then, a human facilitator selects, from a set of possible closed-options, the utterance that semantically resembles the original phrase the most, prompting the appropriate response from the patient. The advantage of this approach is that it ease the development burden of what appears to learners as an \textit{open-option} VP, with the clear disadvantage of preventing its use as a self-learning and self-evaluation tool.

\textbf{Feedback}. With the term \emph{feedback} we refer to any form of instructional scaffolding enclosed in the simulation itself (i.e., we exclude any feedback external to the simulation, such as post-simulation debrief and reflection sessions with mentors and peers).
Feedback can be given in many different forms, from explicit messages to discoveries made, questions answered, and visual representations of the current VP state. 

While researchers recognize the relevance of immediate and after-action feedback as an essential feature in communication-based VPs (\cite{adefila2020students,jacklin2018improving,marei2018use,peddle2019exploring,quail2016student}), 
it is surprising that six of the VPs surveyed (Banszki \cite{banszki2018clinical,quail2016student}, Maicher \cite{maicher2017developing}, Marei \cite{marei2018use}, Szilas \cite{szilas2019virtual}, Washburn \cite{washburn2020virtual}, CESTOLVRClinic \cite{sapkaroski2018implementation}) do not embed any type of built-in feedback. 
The remaining works take different approaches. Four VPs (Dupuy \cite{dupuy2019virtual}, Communicate! \cite{jeuring2015communicate}, VSPR \cite{peddle2019exploring,peddle2019development}, Zlotos \cite{zlotos2016scenario}) offer the possibility, after the simulation is completed, to \textit{replay} some of its parts and analyze the outcome of different choices. 
Three simulations (At-Risk in Primary Care \cite{albright2018using}, Suicide Prevention \cite{o2019suicide},  Schoenthaler \cite{schoenthaler2017simulated}) include a \emph{virtual instructor}, i.e., a virtual tutor that gives advice or feedback based on the user's choices.
MPathic-VR \cite{guetterman2019medical,kron2017using} employs a \emph{multiple session structure}, where the first run acts as a learning phase, concluded by an automated and complete feedback provided by the system, whereas the second run (set in the same scenario) serves as an evaluation phase. This type of structure appears to be highly appreciated by students since they can immediately put into practice what they learned during the first phase, taking into account the feedback received.
Two VPs (At-Risk in Primary Care \cite{albright2018using}, Schoenthaler \cite{schoenthaler2017simulated}) feature \emph{quantitative emotional feedback} in the form of an on-screen trust meter that indicates users how effective their communication choices were at building a relation with the patient. Two VPs (Jacklin \cite{jacklin2019virtual,jacklin2018improving}, Richardson \cite{richardson2019virtual}) offer learners a \emph{personalized qualitative feedback} at the end of the simulation. CynthiaYoungVP \cite{foster2016using} uses an hybrid approach between automated and human feedback. At the end of the simulation, the students can access a web page containing \emph{empathy feedback} and scores for each response given, where scores are manually assigned by human experts. 
The approach adopted in NERVE \cite{hirumi2016advancing,hirumi2016advancingPart2,kleinsmith2015understanding} is to inform learners about the number of \emph{clinical discoveries} and empathic responses available, thus providing inexperienced users with useful guidelines on how to proceed with the conversation. Although it has not been  implemented yet, the work in  \cite{hirumi2016advancingPart2} put forth the proposition of providing \quotes{cumulative feedback} on how users are developing their skills across multiple patient scenarios. Authors suggest that this feature could both encourage repeated use of the system and act as a motivator for performance improvement. 
Finally, there is a number of VPs (HOLLIE \cite{adefila2020students}, Dupuy \cite{dupuy2019virtual}, Communicate! \cite{jeuring2015communicate}, MPathic-VR \cite{guetterman2019medical,kron2017using}, Schoenthaler \cite{schoenthaler2017simulated}, UT-Time Portal \cite{zielke2016beyond,zielke2016using}) that leverage \emph{game elements} as feedback. Since the introduction of game elements is a relevant design feature, it is discussed in detail in the  following subsection.

\textbf{Gamification}. The idea of introducing game mechanics in any learning experience is to make them more enjoyable and engaging.  Researchers and practitioners recognize that game mechanics contribute to making the learning experience more effective, fostering self-improvement and healthy competition between peers \cite{benedict2013promotion,festinger1954theory}. 
The mechanics used the most in \quotes{gamified} experiences are \emph{scores}, \emph{badges} and \emph{leaderboards}. 
Scores are a quantitative and immediate form of feedback that acts as an extrinsic motivator to foster users to improve their performance. \emph{Scoring systems} can be found in Dupuy \cite{dupuy2019virtual}, Communicate! \cite{jeuring2015communicate}, MPathic-VR \cite{guetterman2019medical,kron2017using}, Schoenthaler \cite{schoenthaler2017simulated} and UT-Time Portal \cite{zielke2016beyond,zielke2016using}. In particular, Communicate! \cite{jeuring2015communicate} and Schoenthaler \cite{schoenthaler2017simulated} provide separate scores for each learning goal (e.g., empathy control, language clarity, and pick up of patient's concerns). Such a feature can help learners to tell the areas in which they are already proficient, distinguishing them from those needing improvement.

Badges are visual representations used in games to prove that the player has reached an intermediate goal on his/her road to mastery. Their purpose is twofold: they are a form of gratification to the learners, and they allow trainees to share achievements with peers and educators. Thus, they also represent an extrinsic motivator for improvement.
In our survey,  the only simulation we found that implements a  \emph{badge system} is UT-Time Portal \cite{zielke2016beyond,zielke2016using}. VSPR \cite{peddle2019exploring,peddle2019development} features a system of certificates issued to users at the end of each learning module which shall be regarded as an \quotes{intrinsic-only} motivator, since there is no overarching structure that enables users to see each others' achievements.

Finally, leaderboards (or rankings) are a primarily extrinsic motivator that leverage competition with peers when they compare their performance to that of others (it should be noted that, for highly competitive individuals, the act of \quotes{climbing the leaderboard} can also be seen as a relevant intrinsic motivator independent of the context).
Surprisingly, despite their demonstrated benefits for learning, in our survey, we found no example of public ranking and leaderboards.

A final note is for HOLLIE \cite{adefila2020students}, which implements the very peculiar idea of a Tamagotchi-style VP the learners have to care for (adequately, at regular intervals and in real-time) over two weeks. Here, the leading game mechanics (constant care over a long period) reproduces quite accurately the daily tasks of a nurse leveraging the innate sense of responsibility in the players.

\subsection{Technical features}
\label{sec:technicalDesign}

This category explores, from a technical perspective, the different solutions that can support (and implement) the choices made in the instructional design, i.e., which are the technical features that enable the accomplishment of the envisioned learning activities. These features include the physical devices required to guarantee the exchange of information between the learner and the system, and the possible communication infrastructure needed to run the simulation.

\textbf{Presentation format}. The surveyed works provide learners with different types of outputs aimed to deliver VP information to the learner and presenting the VP itself. A first rough subdivision is between \textit{text-based} and \textit{graphic} representations. In screen-based text simulators, the VP is presented mainly in the form of a collection of text and structured data, with the possible inclusion of images portraying a static patient or exam results. However, the lack of a graphic component capable of displaying a patient that can express emotions as the simulation unfolds (and, consequently, change posture and facial expressions according to its current  state) is one of the main limitations of these approaches. Therefore, researchers started extending text-based simulations into learning activities with a relevant graphic component. 

All the VPs surveyed in this work fall in the \emph{graphic} category, which can be further classified in \emph{image}, \emph{video} and \emph{3D}. VPs in the \emph{image} subclass are presented through a series of static images (either photographs or drawings), such as HOLLIE \cite{adefila2020students} and Marei \cite{marei2018use}.  Some VPs present their case using \emph{video}, either in the form of live footage (Suicide Prevention \cite{o2019suicide}, VSPR \cite{peddle2019exploring,peddle2019development}) or as a computer-generated offline video (Cynthia Young VP \cite{foster2016using}). However, a clear limitation of this approach is its lack of flexibility, since the actor video cannot be re-purposed to portray a different clinical case. 

The majority of surveyed simulations fall in the \emph{3D} subclass and present the patient and the environment as 3D models rendered in real-time.  Their main advantage is that tweaking and expanding a simulation using 3D characters can be done in a much more modular fashion than with \emph{image} and \emph{video}-based VPs. Another advantage is that the sense of immersion and presence are greater than those that can be delivered by \emph{image} and \emph{video}-based VPs. 

Most of the 3D approaches rely on standard \emph{desktop VR} (DVR) settings (10, namely AtRiskInPrimaryCare \cite{albright2018using}, MPathic-VR \cite{guetterman2019medical,kron2017using}, NERVE \cite{hirumi2016advancingPart2,hirumi2016advancing,kleinsmith2015understanding}, Jacklin \cite{jacklin2019virtual,jacklin2018improving}, Communicate! \cite{jeuring2015communicate}, Richardson \cite{richardson2019virtual}, Schoenthaler \cite{schoenthaler2017simulated}, Szilas \cite{szilas2019virtual}, UTTimePortal \cite{zielke2016beyond,zielke2016using}, Zlotos \cite{zlotos2016scenario}). However, since trying to maximize the feeling of immersion and presence is extremely relevant for engaging learners and helping them achieve the expected learning outcomes, some works integrate (partially or fully) immersive technologies. Four of them (Dupuy \cite{dupuy2019virtual}, Banszki \cite{banszki2018clinical,quail2016student}, Maicher \cite{maicher2017developing}, Washburn \cite{washburn2020virtual}) take advantage of \emph{large volume displays}  to portray a life-sized and more natural interaction with the patient, and CESTOL VR Clinic \cite{sapkaroski2018implementation} uses an HMD for the same purpose. In Ochs \cite{ochs2019training}, three different setups (DVR, immersive VR with an HMD, and immersive VR in a CAVE) are compared to analyze their effect on the sense of presence. The outcome of this experiment demonstrates that immersive environments improve the sense of presence and perception of the VP, with the CAVE scoring slightly better than the HMD.
It should be noted, however, that while \emph{immersive VR} (IVR) offers a higher degree of immersion and presence over DVR, there are still accessibility issues that limit its use, in particular when the VP is intended for self-learning and self-training.

\textbf{Input Interface}. This category describes the input methods through which the user influences the unfolding of the VP simulation.
In the case of \emph{typed} interfaces, user's textual intents are entered by typing on a keyboard or selecting an item in a predefined list of choices.
\emph{Voice-controlled} simulations use natural language, which is then parsed into text through a speech-to-text module, usually offered by external Natural Language Processing (NLP) APIs \cite{foster2016using,maicher2017developing}. 
Finally, the integration within the simulation of Natural User Interfaces (NUI) allows to influence the VP state evolution through additional \textit{non-verbal} input cues such as eye contact, distance, facial expression, gestures and body posture, which can be captured with cameras and other hardware. 

Among the analyzed VPs, 14 feature a \emph{typed}-only input, only five are \emph{voice-controlled} (Banszki \cite{banszki2018clinical,quail2016student}, Dupuy \cite{dupuy2019virtual}, MPathic-VR \cite{guetterman2019medical,kron2017using}, Ochs \cite{ochs2019training}, CESTOL VR Clinic \cite{sapkaroski2018implementation}), Maicher \cite{maicher2017developing} has both options, and Washburn \cite{washburn2020virtual} can be considered as a hybrid solution since a human facilitator transcribes the spoken commands through a \emph{typed} interface. 
One of the reasons behind the limited use of voice controls is the fear, expressed by some Authors \cite{maicher2017developing,ochs2019training}, that  NLP systems may be technically hard to implement and prone to wrong transcriptions, which may lead to misunderstandings or unrecognized utterances, break the sense of immersion and cause frustration in the user \cite{bloodworth2010initial}.
This is why some Authors (e.g., Banszki \cite{banszki2018clinical,quail2016student}, Ochs \cite{ochs2019training}, and Washburn \cite{washburn2020virtual}) decided to have a human facilitator taking over the function of the NLP module. 
 Moreover, a VP featuring only voice controls cannot be used by learners with speech impairments \cite{maicher2017developing}. However, it should be stressed that, nowadays, speech-to-text APIs have become widely available, and their quality keeps improving; thus, problems related to imprecise transcriptions should be less and less daunting in the coming years. As for the impaired people, a smart solution to achieve maximum flexibility and accessibility could be to let the users  choose between  \emph{typed} and \emph{voice-controlled} interfaces freely. It should also be noted that IVR environments favour the use of \emph{voice-controlled} interfaces over alternative solutions such as virtual keyboards or situation-specific control boards  
 \cite{sapkaroski2018implementation}, which are likely to break the sense of immersion and presence and are often cumbersome to use.

Among the analyzed VPs, five of them support also \emph{non-verbal} input, by either leveraging NUI-based approaches (e.g., using RGBD sensors, like in MPathic-VR \cite{guetterman2019medical,kron2017using} and Maicher \cite{maicher2017developing}, or standard RGB cameras, like in Dupuy \cite{dupuy2019virtual}) or having a human controller that observes the user interacting with the VP and updates the VP's response accordingly in terms of gestures and facial expressions (like in Banszki \cite{banszki2018clinical,quail2016student}). 
However, apart from Banszki \cite{banszki2018clinical,quail2016student}, it appears that this information is largely underutilized to influence the VP's behavior. In Dupuy \cite{dupuy2019virtual}, the users' facial expressions are detected to merely assess their emotional state at the end of the simulation. In Maicher \cite{maicher2017developing}, the tracked user position is simply used to adjust the agent's gaze, and there is no specific mention of the way the simulation exploits gestures. Finally, in  MPathic-VR \cite{guetterman2019medical,kron2017using}, instead of continuously capturing \emph{non-verbal} communication, learners are forced to assume specific expressions and poses when prompted by the system. In summary, the above discussion highlights that sounder ways of using \emph{non-verbal} inputs are sorely needed in this particular research field.

\textbf{Distribution}. One relevant technological parameter of the VP simulation is the way the application is distributed (and how learners can access it). In principle, there are two main options. The first option is to deploy the VP as a \emph{web-based} application that can be accessed over the Internet. Such a simulation often runs inside a web browser (which makes it device-independent), and generally requires a low amount of computational resources. This flexibility can also foster self-learning (since simulations can be accessed at places and times convenient for the learner) and helps reduce costs (since learning can be carried out online). However, since \emph{web-based} applications are required to be portable on many devices (including mobile ones), they generally sacrifice technical characteristics and computational complexity in favour of accessibility. On the other hand, \emph{standalone} applications are deployed locally on a computer or workstation. These simulations can implement more advanced and complex features since they can leverage the full computational power of a dedicated machine, and integrate external devices or sensors (such as high-quality cameras and microphones).

In our survey, we found a total of ten \emph{web-based} and eight \emph{standalone} applications; in two cases, this information was undisclosed in the paper, whereas in one case (Maicher \cite{maicher2017developing}), the VP was deployed in both variants. This latter work is interesting since it shows how a \emph{standalone} version can trade off some of the flexibility of the web one with a broad array of  features (such as voice control and gesture/posture detection). 
The Authors observed that students were significantly more engaged with the \emph{standalone} VP, whereas in the web version they had to focus on typing and reading, which make them be less prone to notice the subtle non-verbal cues manifested by the patient.


Nonetheless, it should be stressed that technology is advancing rapidly, and personal devices come equipped with ever better microphones, cameras and computational power, which can reduce the technological gap between (desktop-only) \emph{standalone} applications and \emph{web-based} ones. Further discussions on this topic are included in Section \ref{sec:openResearch}.

\subsection{Effectiveness of design elements}
\label{sec:effectiveness}
The general effectiveness of VPs on developing communication skills has been discussed by several Authors \cite{lee2020effective,peddle2016virtual,richardson2019virtualreview}. A common complaint in VP-related literature is the lack of a standardized terminology that, coupled with a considerable heterogeneity in study design, makes the retrieval and evaluation of relevant works a troublesome task. Despite this situation, both \cite{lee2020effective} and \cite{peddle2016virtual} concluded that, when appropriately contextualized in a well thought out educational context, VPs are indeed useful for developing, practising and building confidence about communication and other skills like, e.g., decision making and teamwork.  

Based on these findings, one possible question arising from our review is if the surveyed papers provide pieces of evidence about the effects on learning outcomes and efficacy of the simulation of the different instructional design elements and the technical features available. 
Unfortunately, the answer is negative. In most of the analyzed works, the Authors reported only users' feedback or comments about a particular element/feature, and a direct comparison between different design choices is missing. The only notable exceptions are three. The first one is represented by \cite{ochs2019training}, in which  different presentation formats were assessed, showing that immersive VR technologies yield superior results when compared to non-immersive ones. The second one concerns the distribution method  \cite{maicher2017developing}. The Authors found that a standalone application can provide a considerably higher level of engagement than its web-based counterpart thanks to the possibility to leverage advanced technical features (voice-controlled input and large volume displays) to increase immersion and focus on the task at hand. The third one compared closed and open-option unfolding designs, highlighting the advantages and disadvantages of each variant \cite{hirumi2016advancing}.


%% file: tables/instructionalTable.tex

\begin{table} [t]
\scriptsize{
\begin{center}
    \caption{Synopsis of the reviewed VPs for each instructional design category}
    \label{table:instructionalDesignTable}
    \begin{tabular}{| p{1.8cm} | p{2cm} | p{8cm} |}
    \hline
        \rowcolor{mygray}
        \multicolumn{3}{|c|}{\textbf{Instructional design}}\\
    \hline
        \rowcolor{lightgray}
        \textbf{Category}  & \textbf{Subcategory} & \textbf{Virtual Patients}\\
    \hline
         \multirow{2}{*}{Structure} & \emph{Narrative} & HOLLIE \cite{adefila2020students}, AtRiskInPrimaryCare \cite{albright2018using}, Dupuy \cite{dupuy2019virtual}, CynthiaYoungVP \cite{foster2016using}, Jacklin \cite{jacklin2019virtual,jacklin2018improving}, MPathic-VR \cite{guetterman2019medical,kron2017using}, Communicate! \cite{jeuring2015communicate}, Marei \cite{marei2018use},  Ochs \cite{ochs2019training}, Szilas \cite{szilas2019virtual} \\
    \cline{2-3}
        & \emph{Narrative + Problem solving} & Banszki \cite{banszki2018clinical,quail2016student}, NERVE
        \cite{hirumi2016advancingPart2,hirumi2016advancing,kleinsmith2015understanding},  Maicher \cite{maicher2017developing}, Suicide Prevention \cite{o2019suicide}, VSPR \cite{peddle2019exploring,peddle2019development}, Richardson \cite{richardson2019virtual}, CESTOLVRClinic \cite{sapkaroski2018implementation}, Schoenthaler \cite{schoenthaler2017simulated},   Washburn \cite{washburn2020virtual},  UTTimePortal \cite{zielke2016beyond,zielke2016using}, Zlotos \cite{zlotos2016scenario}\\
    \hline
         \multirow{3}{*}{Unfolding} & \emph{Closed-option} & HOLLIE \cite{adefila2020students}, AtRiskInPrimaryCare \cite{albright2018using}, Dupuy \cite{dupuy2019virtual}, MPathic-VR \cite{guetterman2019medical,kron2017using}, NERVE
        \cite{hirumi2016advancingPart2,hirumi2016advancing,kleinsmith2015understanding}, Jacklin \cite{jacklin2019virtual,jacklin2018improving}, Communicate! \cite{jeuring2015communicate}, Marei \cite{marei2018use}, Suicide Prevention \cite{o2019suicide}, VSPR \cite{peddle2019exploring,peddle2019development}, Richardson \cite{richardson2019virtual}, CESTOLVRClinic \cite{sapkaroski2018implementation}, Schoenthaler \cite{schoenthaler2017simulated}, Szilas \cite{szilas2019virtual}, UTTimePortal \cite{zielke2016beyond,zielke2016using}, Zlotos \cite{zlotos2016scenario}, \\
    \cline{2-3}
        & \emph{Open-option} & Banszki \cite{banszki2018clinical,quail2016student}, CynthiaYoungVP\cite{foster2016using}, NERVE
        \cite{hirumi2016advancingPart2,hirumi2016advancing,kleinsmith2015understanding}, Maicher \cite{maicher2017developing}, Washburn \cite{washburn2020virtual} \\
    \cline{2-3}
        & \emph{Hybrid} & Ochs \cite{ochs2019training}  \\
    \hline
        \multirow{8}{*}{Feedback} & \emph{Replay feature} & Dupuy \cite{dupuy2019virtual}, Communicate! \cite{jeuring2015communicate}, Ochs \cite{ochs2019training}, VSPR \cite{peddle2019exploring,peddle2019development}, Zlotos \cite{zlotos2016scenario}\\
    \cline{2-3}
        & \emph{Virtual instructor} & At-Risk in Primary Care \cite{albright2018using}, Suicide Prevention \cite{o2019suicide}, Schoenthaler \cite{schoenthaler2017simulated}\\
    \cline{2-3}
        & \emph{Multiple session structure} & MPathic-VR \cite{guetterman2019medical,kron2017using}  \\
    \cline{2-3}
        & \emph{Quantitative emotional feedback} & At-Risk in Primary Care \cite{albright2018using}, Schoenthaler \cite{schoenthaler2017simulated}\\
    \cline{2-3}
        & \emph{Qualitative personalized post-feedback} & Jacklin \cite{jacklin2019virtual,jacklin2018improving}, Richardson \cite{richardson2019virtual}\\
    \cline{2-3}
        & \emph{Empathy feedback} & CynthiaYoungVP \cite{foster2016using}\\
    \cline{2-3}
        & \emph{Clinical discoveries available} & NERVE
        \cite{hirumi2016advancingPart2,hirumi2016advancing,kleinsmith2015understanding}\\
    \cline{2-3}
        & \emph{Game elements} & Dupuy \cite{dupuy2019virtual}, MPathic-VR \cite{guetterman2019medical,kron2017using}, Communicate! \cite{jeuring2015communicate}, Schoenthaler \cite{schoenthaler2017simulated}, UT-Time Portal \cite{zielke2016beyond,zielke2016using}\\
    \hline
        \multirow{3}{*}{Gamification} & \emph{Scoring system} & Dupuy \cite{dupuy2019virtual}, MPathic-VR \cite{guetterman2019medical,kron2017using}, Communicate! \cite{jeuring2015communicate}, Schoenthaler \cite{schoenthaler2017simulated}, UT-Time Portal \cite{zielke2016beyond,zielke2016using}\\
    \cline{2-3}
        & \emph{Badge system} & UTTimePortal \cite{zielke2016beyond,zielke2016using}  \\
    \cline{2-3}
        & \emph{Countdown timed events} & HOLLIE \cite{adefila2020students}\\
    \hline
     \end{tabular}
\end{center}
}
\end{table}
\normalsize

%% file: tables/technicalTable.tex

\begin{table} [t]
\scriptsize{
\begin{center}
    \caption{Synopsis of the reviewed VPs for each technical design category}
    \label{table:technicalDesignTable}
    \begin{tabular}{| p{1.8cm} | p{2cm} | p{8cm} |}
    \hline
        \rowcolor{mygray}
        \multicolumn{3}{|c|}{\textbf{Technical Design}}\\
    \hline
        \rowcolor{lightgray}
        \textbf{Category}  & \textbf{Subcategory} & \textbf{Virtual Patients}\\
    \hline
        \multirow{5}{1.8cm}{Presentation format} & \emph{Image} & HOLLIE \cite{adefila2020students},  Marei \cite{marei2018use}\\
    \cline{2-3}
        & \emph{Video} & CynthiaYoungVP \cite{foster2016using}, Suicide Prevention \cite{o2019suicide}, VSPR \cite{peddle2019exploring,peddle2019development}\\
    \cline{2-3}
        & \emph{Desktop VR} & AtRiskInPrimaryCare \cite{albright2018using}, MPathic-VR \cite{guetterman2019medical,kron2017using}, NERVE
        \cite{hirumi2016advancingPart2,hirumi2016advancing,kleinsmith2015understanding}, Jacklin \cite{jacklin2019virtual,jacklin2018improving}, 
        Communicate! \cite{jeuring2015communicate}, Richardson \cite{richardson2019virtual}, Schoenthaler \cite{schoenthaler2017simulated}, Szilas \cite{szilas2019virtual}, UTTimePortal \cite{zielke2016beyond,zielke2016using}, Zlotos \cite{zlotos2016scenario}\\
    \cline{2-3}
        & \emph{Large volume display} & Dupuy \cite{dupuy2019virtual}, Banszki \cite{banszki2018clinical,quail2016student}, Maicher \cite{maicher2017developing}, Washburn \cite{washburn2020virtual}\\
    \cline{2-3}
        & \emph{Immersive VR} & Ochs \cite{ochs2019training}, CESTOLVRClinic \cite{sapkaroski2018implementation}\\
    \hline
        \multirow{4}{1.8cm}{Input interface} & \emph{Typed} & HOLLIE \cite{adefila2020students}, AtRiskInPrimaryCare \cite{albright2018using}, CynthiaYoungVP \cite{foster2016using}, NERVE
        \cite{hirumi2016advancingPart2,hirumi2016advancing,kleinsmith2015understanding}, Jacklin \cite{jacklin2019virtual,jacklin2018improving}, Communicate! \cite{jeuring2015communicate}, Maicher \cite{maicher2017developing}, Marei \cite{marei2018use}, Suicide Prevention \cite{o2019suicide}, VSPR \cite{peddle2019exploring,peddle2019development}, Richardson \cite{richardson2019virtual}, CESTOLVRClinic \cite{sapkaroski2018implementation}, Schoenthaler \cite{schoenthaler2017simulated}, Szilas \cite{szilas2019virtual}, UTTimePortal \cite{zielke2016beyond,zielke2016using}, Zlotos \cite{zlotos2016scenario}\\
    \cline{2-3}
        & \emph{Voice-controlled} & Banszki \cite{banszki2018clinical,quail2016student}, Dupuy \cite{dupuy2019virtual}, MPathic-VR \cite{guetterman2019medical,kron2017using}, Maicher \cite{maicher2017developing}, Ochs \cite{ochs2019training}, CESTOLVRClinic \cite{sapkaroski2018implementation}\\
    \cline{2-3}
        & \emph{Hybrid} & Washburn \cite{washburn2020virtual} \\
    \cline{2-3}
        & \emph{Non-verbal} & Banszki \cite{banszki2018clinical,quail2016student}, Dupuy \cite{dupuy2019virtual}, MPathic-VR \cite{guetterman2019medical,kron2017using}, Maicher \cite{maicher2017developing}, CESTOLVRClinic \cite{sapkaroski2018implementation}\\
    \hline
        \multirow{3}{*}{Distribution} & \emph{Standalone} & Banszki \cite{banszki2018clinical,quail2016student}, Dupuy \cite{dupuy2019virtual}, MPathic-VR \cite{guetterman2019medical,kron2017using}, Maicher \cite{maicher2017developing}, Marei \cite{marei2018use}, Ochs \cite{ochs2019training}, CESTOLVRClinic \cite{sapkaroski2018implementation}, Szilas \cite{szilas2019virtual}, Washburn \cite{washburn2020virtual}\\
    \cline{2-3}
        & \emph{Web-based} & HOLLIE \cite{adefila2020students}, AtRiskInPrimaryCare \cite{albright2018using}, CynthiaYoungVP \cite{foster2016using}, NERVE
        \cite{hirumi2016advancingPart2,hirumi2016advancing,kleinsmith2015understanding}, Jacklin \cite{jacklin2019virtual,jacklin2018improving}, Maicher \cite{maicher2017developing}, Suicide Prevention \cite{o2019suicide}, VSPR \cite{peddle2019exploring,peddle2019development}, Richardson \cite{richardson2019virtual}, UTTimePortal \cite{zielke2016beyond,zielke2016using}, Zlotos \cite{zlotos2016scenario}\\
    \cline{2-3}
         & \emph{Undisclosed} &  Communicate! \cite{jeuring2015communicate}, Schoenthaler \cite{schoenthaler2017simulated} \\
    \hline
     \end{tabular}
\end{center}
}
\end{table}
\normalsize

%% file: openResearch.tex
\section{Open areas of research}
\label{sec:openResearch}

The surveyed papers  show that, despite exciting results obtained, fully understanding how to develop effective VPs for patient-doctor communication training requires further work. Reasons are related to the fact that either the technological components have not been fully explored yet or results are still inadequate to fully assess the effectiveness of different design approaches. Thus, in this section, we briefly discuss some open problems and present areas requiring further research.

\textbf{Assessment of design elements.}
As discussed in Section \ref{sec:effectiveness}, the current literature lacks a thorough evaluation of the effectiveness of alternative designs. This observation highlights the fact that further work has to be done to develop a better understanding of instructional elements and technical features that VP simulations can offer in order to achieve the desired learning outcomes.

\textbf{Scope.} Another comment can be made on the specific communication learning context. While several core skill domains jointly contribute to a patient's health and satisfaction (like relationship building, information gathering, patient education, shared decision making and breaking bad news \cite{riedl2017influence}), most of the surveyed VP simulations  focus only on one specific domain. This observation highlights the need to develop novel approaches capable of addressing simultaneously the multiple communication challenges one has to face when interacting with a real patient, thus helping to improve the overall learner's communication skills.

\textbf{Authoring tools.}
Implementing VPs is a cumbersome and complicated process, which requires taking into account several different elements (NLP, emotion modelling, affective computing, 3D animations, etc.), which, in turn, involve specific technological and technical skills. Usually, the development of a VP is a cyclical process of research, refinement and validation with experts that can take a considerable amount of time \cite{rossen2009human}. Thus, there is the need to develop simple (and effective) authoring tools that can allow developers to support clinical educators in the rapid design, prototyping and deploying of VPs in a variety of use cases. 
Examples of authoring tools for narrative-style VPs with 3D graphics are very scarce in the literature. The work presented in \cite{jeuring2015communicate} integrates a scenario builder that allows clinical educators to design the unfolding of their cases. This authoring tool exploits a domain reasoner where the response of the virtual agent is determined not only by the previous dialogue that the user chose, but also by other parameters like the agent's current emotional state. However, this tool lacks the possibility to customize the virtual environment or the VP's aesthetics.
The NERVE VP \cite{hirumi2016advancingPart2,hirumi2016advancing,kleinsmith2015understanding} is built upon the Virtual People Factory \cite{rossen2009human}, a web application that enables the users to build conversational models using an un-annotated corpus retrieval approach based on keyword matching. 
Another interesting example is SIDNIE (Scaffolded Interviews Developed by Nurses in Education \cite{dukes2016participatory}). This tool allows clinical educators to edit the patient's medical status, dialogue options and physical appearance. However, to our knowledge, SIDNIE has not been deployed in any publicly available form, and  appears to be aimed exclusively at nurse training scenarios.

In other application areas (such as building clinical skills and problem-solving abilities), the extensive use of tools such as DecisionSim, OpenLabyrinth and Web-SP 
\cite{doloca2015comparative} is a clear demonstration of the fact that an easy-to-use authoring tool is a determinant factor for the success of a VP application. However, compared to these areas, the specific context of patient-doctor communication training involves more complex systems, with 3D visuals and branched narratives offering a more realistic interaction, which makes the development of authoring tools in this area much more challenging \cite{talbot2012sorting}.\par

\textbf{Emerging web technologies.}
In the previous sections, we highlighted that personal devices are coming with better and better hardware and computational power, thus helping to narrow the gap between standalone and web-based applications. Another contribution will inevitably come from recent advances in web-based technology, like, e.g., WebXR\footnote{\url{https://www.w3.org/TR/webxr/}}. WebXR is a device-independent framework that allows users to develop and share VR and AR applications over the Internet, with considerable support for different hardware and web browsers. In addition, game-streaming platforms such as Google Stadia\footnote{\url{https://stadia.google.com/}} are a very promising workaround for the limited computational capabilities of personal devices. With these platforms, the bulk of the computation is processed on the server side, then the pre-rendered output is streamed to the final user's device. The implementation of such technological solutions in the immediate future will enable the applications to combine the accessibility of current web-based software with the computational complexity of standalone applications run on a dedicated machine. \par


\textbf{Multiple virtual humans.}
Interacting with a relative or another health care provider are considered crucial aspects of a clinician's communication skills \cite{hallin2011effects,kee2018communication}. However, VP simulations usually include only two actors: the learner (possibly represented by an avatar) and a unique Non-Playable Character (NPC), i.e, a virtual human not controlled by the trainee that represents the patient. The only two examples that include more than one NPC besides the patient are the Medical Interview Episode of the UTTimePortal \cite{zielke2016beyond,zielke2016using} (which incorporates a patient and a caregiver), and MPathic-VR \cite{guetterman2019medical,kron2017using} (which includes a patient's relative and a nurse). 
Beyond this observation, we should also note that another interesting future development (still untouched in the field of VPs for patient-doctor communication skills, to the best of our knowledge) could be to provide the possibility of interacting (within the simulation) with other human-controlled avatars, in a way similar to that proposed by approaches focused on inter-professional communication in emergency medical situations \cite{anbro2020using}.

\textbf{Immersive VR and AR.}
There is a general understanding among researchers that increasing the level of immersion and realism of the simulations (e.g., using large volume displays, HMDs, spatialized 3D audio, higher fidelity graphics and animations) leads to more believable human-computer interactions \cite{chuah2013exploring,johnsen2008evaluation}, which in turns help improve the users' communication and empathic skills  \cite{ochs2019training,zielke2017developing} and, ultimatley, the learning outcomes in general \cite{limniou2008full}. 
However, surprisingly, the use of IVR technologies in this specific context appears to be quite limited. Only two VPs out of \totalVPs, i.e., Ochs \cite{ochs2019training} and CESTOL VR Clinic \cite{sapkaroski2018implementation}, mention the use of IVR, and AR appears to be completely unexplored.  The primary obstacles to the adoption of IVR or AR in VP simulations seem to be the complexity, challenges and costs of development steps \cite{zielke2017developing}.  

Fortunately, things are going to change rapidly. In recent years, the availability and quality of VR devices have increased considerably, and their cost has decreased dramatically.  These factors contribute (together with the availability of high-end development platforms such as Unity or Unreal engine) to reducing overall costs and efforts for developing IVR and AR applications. Furthermore, IVR offers currently a truly immersive, unbroken environment that can shift the cognitive load directed on imagining oneself \quotes{being there} in VR towards solving the task at hand. In turn, higher immersion and visual fidelity can have positive effects on learning \cite{coulter2007effect}, \cite{huerta2012measuring}. Thus, we expect that, soon, VR and AR will contribute to improving the state of the art in this research field.

\textbf{Fully-fledged non-verbal input.}
In our opinion, this is a major lack in current designs. The unfolding of the simulation's narrative should be dictated (in tandem) by both user's verbal and non-verbal behaviours. To this end, developers of future VPs should attempt to fully leverage non-verbal cues as a factor that actively influences the state of the agent. For instance, the same utterance should have a different outcome if the user maintains eye contact with the patient, looks in another direction, and is fidgeting or exhibiting an incoherent facial expression.
The extraction of para-linguistic factors such as tone of voice, loudness, inflection, rhythm, and pitch can provide information about the actual emotional states of the other peer in the communication. Prosody must be addressed with great attention since it is one of the main ways to express empathy and can have a considerable impact in increasing patient satisfaction \cite{kee2018communication}. Thus, computational mechanisms capable of extracting these variables from the analysis of the user's voice are sorely needed. 
The same para-linguistic factors should be also available to modulate the VP response according to its emotional states. In fact, one of the problems with present text-to-speech libraries is that they pronounce everything with the same tone, which makes it impossible to communicate feelings through voice.

%% file: conclusion.tex
\section{Conclusion}
\label{sec:conclusion}

In our research we found many different examples of VPs focused on provider-patient communication and various approaches to their design. However, we feel that there is not a single VP that realizes the full potential of this learning tool. Some research areas still need to be explored further. The broad range of educational use cases in healthcare suggests that VP applications should be as modular and adaptable as possible. Effective and user-friendly authoring tools are very rarely implemented while being, in our opinion, a crucial feature to ensure the adoption of a VP simulation by clinical educators. 
The use of technologies such as VR, AR, and advanced NLP also needs to be explored more in depth, as they may give VP simulations the edge they need to be effectively used in daily practice. We also feel that recent developments in web-based technologies will also reduce those compromises between accessibility and advanced technical possibilities that today are still required in many situations.
